\documentclass[aps,floatfix,prl,superscriptaddress,groupedaddress]{revtex4}	

\usepackage{graphicx,amsmath,bm,floatrow}

\begin{document}

\title{Green formulation for studying electromagnetic scattering from graphene--coated wires of arbitrary section}





\author{{Claudio Valencia$^{1}$, M\'aximo A. Riso$^2$, Mauro Cuevas$^{3}$, and Ricardo A. Depine$^{2,*}$} \\
{\small \em $^1$ Facultad de Ciencias, Universidad Aut\'onoma de Baja California (UABC), Ensenada, BC 22860, M\'exico\\
$^2$Grupo de Electromagnetismo Aplicado, Departamento de F\'{\i}sica, FCEN, Universidad de Buenos Aires and IFIBA, Consejo Nacional de Investigaciones Cient\'{\i}ficas y T\'{e}cnicas (CONICET), Ciudad Universitaria, Pabell\'{o}n I, C1428EHA, Buenos Aires, Argentina \\
$^3$ Facultad de Ingenier\'ia y Tecnolog\'ia Inform\'atica, Universidad de Belgrano, Villanueva 1324, C1426BMJ, Buenos Aires, Argentina and Consejo Nacional de Investigaciones Cient\'{\i}ficas y T\'{e}cnicas (CONICET)\\}
$^*$email: rdep@df.uba.ar}



\begin{abstract}
We present a rigorous electromagnetic method based on Green's second identity for studying the plasmonic response 
of graphene--coated wires of arbitrary shape. The wire is illuminated perpendicular to its axis by a 
	monochromatic electromagnetic wave and the wire substrate is homogeneous and isotropic. 
The field is expressed everywhere in terms of two  unknown source functions evaluated on the graphene coating which can be obtained from the numerical solution of a coupled pair of inhomogeneous integral equations. 
To assess the validity of the Green formulation, the scattering and absorption efficiencies obtained numerically in the particular case of circular wires are compared with those obtained from the  multipolar Mie theory. 
An excellent agreement is observed in this particular case, both for metallic and dielectric substrates. 
To explore the effects that the break of the rotational symmetry of the wire section introduces in the plasmonic features of the scattering and absorption response, the Green formulation is applied to the case of graphene-coated wires of elliptical section. 
As might be expected from symmetry arguments, we find a two-dimensional anisotropy in the angular optical 
response of the wire, particularly evident in the frequency splitting of multipolar plasmonic resonances. 
%
The comparison between the spectral position of the enhancements in the scattering and absorption efficiency 
spectra for low--eccentricity elliptical and circular wires allows us to guess the multipolar order of each plasmonic resonance. 
%
We present calculations of the near field distribution for different frequencies which explicitly reveal the 
multipolar order of the plasmonic resonances. They also confirm the previous guess and serve as a further test 
on the validity of the Green formulation. 

%
\end{abstract}


\maketitle

\section{Introduction}\label{intro}
Due to its particular electronic band structure, the atom-thick form of carbon known as graphene exhibits unique electronic and optical properties that have attracted tremendous attention in recent years \cite{geim1}. 
Several applications from terahertz (THz) to visible frequencies, including solar cells, touch screens, photodetectors, light-emitting devices and ultrafast lasers, are clear evidence of the rise of graphene in photonics and optoelectronics \cite{bonaccorso1,bonaccorso2}. The strength of the mutual interaction between graphene and electromagnetic radiation plays a key role in a great majority of these applications. However, a single sheet of homogeneous graphene exhibits an optical absorbance of $\sim 2.3$\% \cite{ssc1}, a value strong enough to detect exfoliated monolayers by visual inspection under an optical microscope but not sufficiently strong as could be desirable in many  photonics and optoelectronics applications. 

The interaction between graphene and electromagnetic radiation can be improved in the presence of surface plasmons, whether by combining a graphene layer with conventional plasmonic nanostructures based on noble metals \cite{conven1,conven2} or by taking advantage of the long-lived, electrically tunable surface plasmons supported by graphene \cite{sp-graf1,sp-graf2}. The first alternative fits well in the visible and near-infrared frequencies where the interband loss becomes large and graphene behaves as a dielectric material, whereas the second alternative fits well in the 
terahertz and infrared regions where doped graphene nanostructures can support surface plasmons \cite{sp-graf1,sp-graf2}. 

Surface plasmons can be roughly divided into two categories: surface plasmon polaritons (SPPs) propagating along  waveguiding structures and localized surface plasmons (LSPs) supported by spatially limited structures, such as scattering particles. 
Since the spatial periodicity associated with a surface plasmon propagating along a graphene monolayer is always less than the spatial periodicity which could be induced by an incident plane wave, plane waves cannot resonantly excite propagating surface plasmons at a flat graphene monolayer. 
In contrast, in bounded geometries, localized surface plasmons (LSPs) can be resonantly excited by plane waves at discrete frequencies that depend on the size and shape of the object to which they are confined.  

The plasmonic properties of graphene wrapped particles have recently attracted the attention of researchers   \cite{esfera2,esfera1,esfera3,cil4,cilindroconico,cilOE,cilindros1,cilindros2,cilindros3,cilindros4}. 
Analytical solutions are only available for particles with a very simple shape, such as spheres 
\cite{esfera2,esfera1,esfera3} or circular cylinders \cite{cilindroconico,cilOE,cilindros1,cilindros2,cilindros3,cilindros4}. 
In these cases, the application of Mie theory leads to multipole coefficients for the scattered field which have essentially the same form as those corresponding to the bare particles, except for additive corrections proportional to the graphene surface conductivity in the numerator and denominator.  
Taking into account the significant progress made in the fabrication of particles with a variety of shapes and dimensions \cite{fabric1} and that wrapping up particles with graphene coatings brings extra freedom for scattering engineering and for improving the interaction between graphene and electromagnetic radiation via surface plasmon mechanisms, the investigation of the scattering characteristics of graphene--coated particles seems a particularly 
interesting and very promising area to explore. 
In the particular case of dielectric, almost transparent particles, graphene coatings introduce tunable plasmons which are absent in the bare particle, whereas for metallic or metallic-like particles, graphene coatings can modify in a controlled manner the LSPs already existing in the bare particle. 

The particular relation between the shape of a graphene particle and its plasmonic characteristics --such as cross-section enhancement, interplay between near- and far--field quantities, plasmon resonance frequencies and linewidths-- can be exploited in many THz applications, including the sensing of small changes in a host-medium refractive index caused by the change in concentration of an analyzed substance \cite{cilindros4} or the design of sub--wavelength metamaterials \cite{C2gribonexp01}.  
In order to realize graphene--wrapped plasmonic particles with tailored properties for specific applications, complete electromagnetic solutions are needed. The purpose of this paper is to present a rigorous, fully retarded method  based on Green's second identity \cite{MMMM,civ2NL} for modeling the scattering characteristics of graphene--coated, homogeneous 
dielectric or metallic two dimensional particles (wires) of rather arbitrary shape. 

The paper is organized as follows. First, in Section \ref{sec:Green} we give exact expressions for the electromagnetic field scattered by a graphene--coated wire illuminated perpendicular to its axis by a p-- or s--polarized monochromatic plane wave. The scattered field is expressed in terms of two  unknown source functions evaluated on the wire surface, 
one related to the total field in the medium of incidence and the other related to its normal derivative. 
To find both surface source functions, a coupled pair of inhomogeneous integral equations must be 
solved numerically. 
In Section \ref{sec:results} the numerical technique is illustrated and validated for wires of circular and elliptical section. 
As a first validation 
we consider metallic and dielectric circular wires and show that in this case the numerical results agree perfectly well with those obtained analytically using Mie's theory \cite{cilindros1,cilindros2,cilindros3}. 
To explore the effects that the departure from circular geometries has on the scattering and extinction cross-sections we then consider graphene--coated wires of elliptical section. 
We present scattering and absorption efficiency spectra showing that, similarly to the plasmonic resonances of bare metallic wires, the break of the rotational symmetry of the wire section introduces a two-dimensional anisotropy in the angular optical response of the wire, particularly evident in a frequency splitting of the strong dipolar plasmonic resonance \cite{martin2,martin3},  but also observable for higher multipolar resonances.  
Using the Green formulation, we show near-field distributions which reveal that 
the splittings correspond to strong localization of the near field at specific positions along the ellipse axes. 
Besides, and as a further test on the validity of the Green formulation, we show that the multipolar order inferred 
for ellipses with low eccentricities from the topology of the near--field distribution for a given resonant frequency  
is in excellent agreement with the multipolar order inferred from  the spectral proximity between the enhancement 
peaks in the scattering efficiency spectra of an elliptical and a similar circular wire. 
Finally, in Section \ref{sec:final} we summarize and discuss the results obtained. The Gaussian system of units is used and an $\exp(-i\omega t)$ time--dependence is implicit throughout the paper, with $\omega$ the angular frequency, $c$ the speed of light in vacuum, $t$ the time, and $i=\sqrt{-1}$. 
	
\section{Green's approach}\label{sec:Green}

\begin{figure}[htbp]
\centering
\includegraphics[width=7cm]{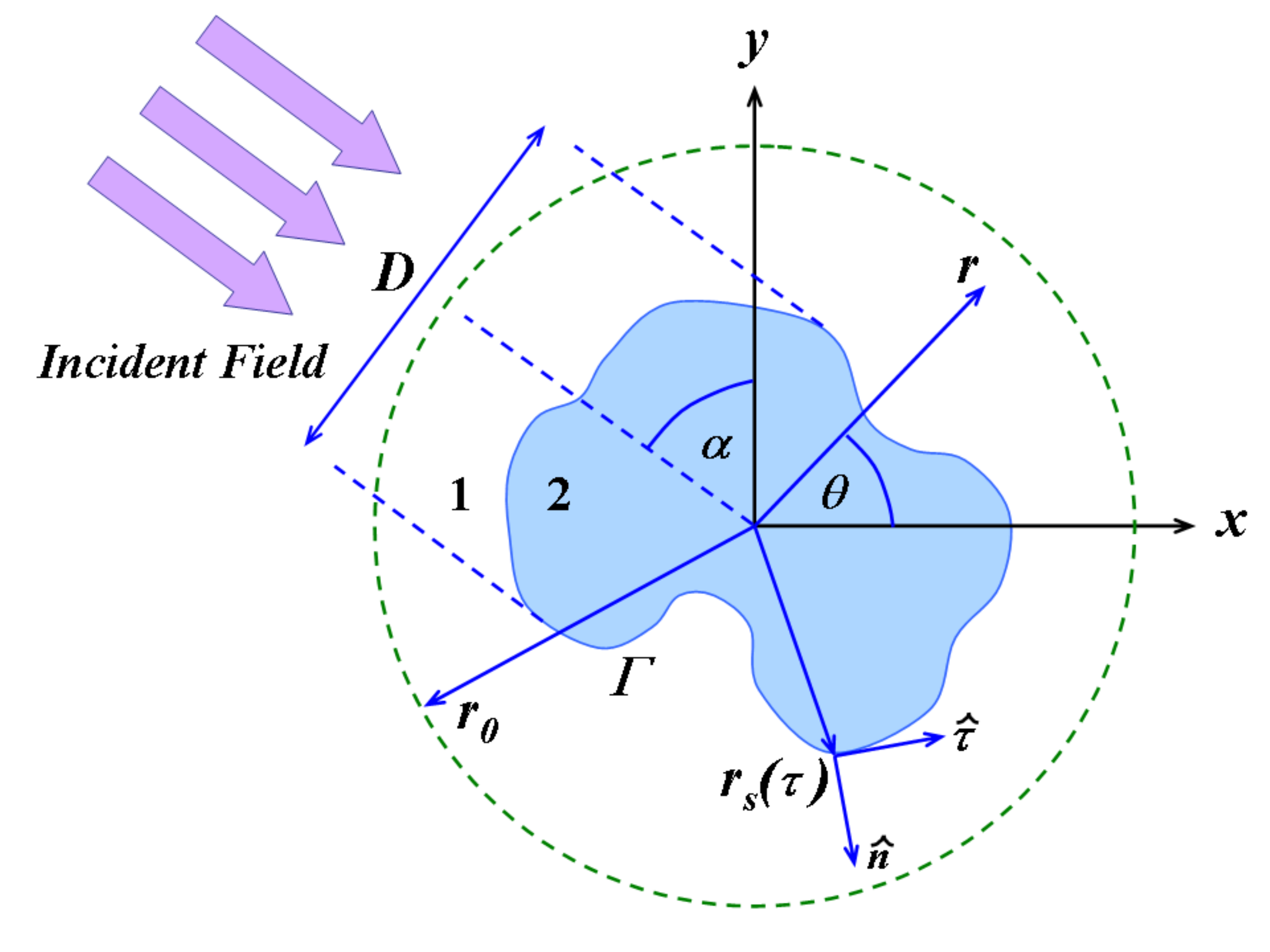}
\caption{Scattering configuration. The main section of the graphene--coated cylinder is defined by the planar curve $\Gamma$. }
\label{fig:esquema}
\end{figure}

We consider a wire in the form of an infinite cylinder whose axis lie along the $\hat z$ axis and whose cross--section is defined by the planar curve $\Gamma$ described by the vector valued function $\textbf{r}_s(\tau) = f(\tau)\hat{x}+g(\tau)\hat{y}$ (see Figure \ref{fig:esquema}). 
The parameter $\tau$ can represent the arc length along the curve or any other convenient parameter, such as the angle $\theta$.
The wire substrate (region 2) is characterized by the electric permittivity $\varepsilon_{2}$ and the magnetic permeability $\mu_{2}$ and is embedded in a transparent medium (region 1) with electric permittivity  $\varepsilon_{1}$ and magnetic permeability $\mu_{1}$. This wire is coated with a graphene monolayer which can be considered as an infinitesimally thin, local and isotropic two-sided layer with frequency--dependent surface conductivity $\sigma(\omega)=\sigma ^{intra}+\sigma ^{inter}$ given by the Kubo formula \cite{kubo1,kubo2}, with the intraband contribution 
$\sigma ^{intra}$ given by 
\begin{equation}
\sigma ^{intra}(\omega) =
\frac{2ie^2T} {\pi\hbar(\omega+i\gamma_c)}
\ln{[2\cosh(\mu_c /2T)]} ,\label{sigma1}
\end{equation}
and the interband contribution $\sigma ^{inter}$ given by 
\begin{equation}
\sigma ^{inter}=
\frac{e^2}{4\hbar} \Big\{\Theta(\hbar \omega-2\mu_c)-\frac{i}{\pi}ln\Big|\frac{\hbar \omega+2\mu_c}{\hbar \omega-2\mu_c}\Big|\Big\} ,\label{sigma2} 
\end{equation}
where $\mu_c$ is the chemical potential (controlled with the help of a gate voltage), $\gamma_c$ the carriers scattering rate, $\Theta(x)$  the Heaviside function, $e$  the electron charge, $k_B$ the Boltzmann constant and $\hbar$ the reduced Planck constant. 
The intraband contribution \eqref{sigma1} dominates for large doping $\mu_c>>k_B\,T$ and is a generalization of the Drude model for the case of arbitrary band structure, whereas the interband contribution \eqref{sigma2}  dominates for large frequencies $\hbar \omega \gtrsim \mu_c$. 

When the wavevector of the incident plane wave 
is perpendicular to the wire axis, the scattering problem can be decomposed into two independent scalar problems: electric field parallel to the main section of the cylindrical surface (p polarization, magnetic field along  $\hat z$) and magnetic field parallel to the main section of the cylindrical surface (s polarization, electric field along  $\hat z$). 
In region $j$ ($j=1,2$) and for each polarization mode we denote by $\psi^{(j)}({\bf r})$ the non-zero component of the total electromagnetic field along the axis of the cylinder, 
evaluated at the observation point ${\bf r}=x\,{\bf \hat x}+ y\,{\bf \hat y}$. 
These field components must satisfy Helmholtz equations in each region 
\begin{equation}
\left(\nabla^2+\Big(\frac{\omega}{c}\Big)^2\varepsilon_j\,\mu_j \right) \psi^{(j)}=0 \,. \,\,\,j=1,2
\label{eq:helmholtz}
\end{equation}
As in the case of uncoated cylinders \cite{MMMM,civ2NL}, the starting point for the derivation of an exact expression for the scattered field is Green's second integral. Using \eqref{eq:helmholtz} in the exterior region and separating the total field in this region into contributions from the incident and scattered field, $\psi^{(1)}(\textbf{r})=\psi^{(1)}_{inc}(\textbf{r})+\psi^{(1)}_{sc}(\textbf{r})$, 
we obtain
\begin{align*}
\psi^{(1)}(\textbf{r})=\psi^{(1)}_{inc}(\textbf{r})
\end{align*}
\begin{align}
+\frac{1}{4\pi}
\int_{\Gamma}\Big( \frac{\partial G^{(1)}(\textbf{r},\textbf{r'})}{\partial \hat{n}'}\psi^{(1)}(\textbf{r'})-G^{(1)}(\textbf{r},\textbf{r'}) \frac{\partial \psi^{(1)}(\textbf{r'})}{\partial \hat{n}'}\Big)dS',
\label{eq:Green2}
\end{align}
where $\textbf{r'}$ is a point on the boundary $\Gamma$ with arc element $dS'$ and $G^{(1)}(\textbf{r},\textbf{r'})$ is the Green function of  \eqref{eq:helmholtz} in the exterior region. Analogously, using Green's second integral and \eqref{eq:helmholtz} in the interior region, we obtain 
\begin{align}
0=
\int_{\Gamma}\Big( \frac{\partial G^{(2)}(\textbf{r},\textbf{r'})}{\partial \hat{n}'}\psi^{(2)}(\textbf{r'})-G^{(2)}(\textbf{r},\textbf{r'}) \frac{\partial \psi^{(2)}(\textbf{r'})}{\partial \hat{n}'}\Big)dS',
\label{eq:Green1}
\end{align}
where $G^{(2)}(\textbf{r},\textbf{r'})$ is the Green function of  \eqref{eq:helmholtz} in the interior region. 
Due to the cylindrical symmetry both Green functions may be expressed in terms of the 
zeroth-order Hankel function of the first kind $H_{0}^{(1)}$, 
\begin{equation}
G^{(j)}({\bf r}|{\bf r}^{\,\,\prime})=i\pi
H_{0}^{(1)}\left( k_{j} |{\bf r}-{\bf r}^{\,\,\prime}|\right) \,,
\label{eq:def_h0}
\end{equation}
with $k_{j}=\frac{\omega}{c}\sqrt{\varepsilon_{j}\mu_{j}}$ ($j=1, 2$). 
By letting the point of observation \textbf{r} approach the surface \textbf{r'} in expressions \eqref{eq:Green2} and \eqref{eq:Green1}, we obtain a pair of integral equations with four unknown functions: the values of the fields $\psi^{(j)}$ and of  their normal derivatives 
$\partial \psi^{(j)} / \partial \hat{n}$, 
j=1,2, at the boundary $\Gamma$. As in the case of uncoated cylinders, the number of unknowns can be reduced to two since the electromagnetic boundary conditions at $\Gamma$  provide two additional relationships between the normal derivatives and the fields at the boundary. 
Taking into account that because of the graphene coating the tangential components of the magnetic field \textbf{H} are no longer continuous across the boundary $\Gamma$ --as they were in the case of uncoated cylinders-- the boundary conditions for our case can be expressed as 
\begin{align}
\frac{1}{\varepsilon_{1}} \frac{\partial \psi^{(1)}}{\partial \hat{n}} 
           = 
\frac{1}{\varepsilon_{2}} \frac{\partial \psi^{(2)}}{\partial \hat{n}} \,\,\,\,\,\,\,\,\,\,\mbox{and}\,\,\,\,\,\,\,\,\,\,
\psi^{(1)}- \psi^{(2)}  =  
\frac{4i\pi}{\omega \varepsilon_{1}}\sigma \frac{\partial \psi^{(1)}}{\partial \hat{n}}  \,, 
\label{eq:contorno-p}
\end{align}
for $p$-polarization, and
\begin{align}
\psi^{(1)}= \psi^{(2)} \,\,\,\,\,\,\mbox{and}\,\,\,\,\,\,
\frac{1}{\mu_{1}} \frac{\partial \psi^{(1)}}{\partial \hat{n}} - 
\frac{1}{\mu_{2}} \frac{\partial \psi^{(2)}}{\partial \hat{n}}      
=-\frac{4i\pi\omega\sigma}{c^2} \psi^{(1)}  \,,
\label{eq:contorno-s}
\end{align}
for $s$-polarization. 
In equations \eqref{eq:contorno-p} and \eqref{eq:contorno-s} the fields and their normal derivatives are evaluated at $\textbf{r}=\textbf{r}_s(\tau)$. 

The boundary conditions allow us to express  $\psi^{(2)}$ and 
$\partial \psi^{(2)}/\partial \hat{n}$ 
in terms of $\psi^{(1)}$ and 
$\partial \psi^{(1)}/\partial \hat{n}$. 
Therefore, eqs \eqref{eq:Green2} and \eqref{eq:Green1} can be rewritten as a set of coupled, inhomogeneous integral equations for the unknown exterior functions $\psi^{(1)}$ and 
$\partial \psi^{(1)}/\partial \hat{n}$.  
To find these functions, the system of integral equations is converted into matrix equations which are then solved numerically. We do this by using a set $[t_{1},...,t_{N}]$ for discretizing the interval where the parameter $\tau$ describing the boundary varies. 
The evaluation of the matrix elements and the treatment of the Hankel functions singularities when the argument is zero follows closely that described in \cite{MMMM,civ2NL}. 
Once the functions $\psi^{(1)}$ and $\partial \psi^{(1)}/\partial \hat{n}$ are known, the scattered field, given by the second term in \eqref{eq:Green2}, can be calculated at every point in the exterior region. Yet another application of Green's second integral in the interior region gives the following expression for the field at every point inside the wire 
\begin{eqnarray}
\centering
\psi^{(2)}({\bf r}) = -\frac{i}{4} \;\int_{\Gamma} \left( k_2\,  
\hat n^{\prime}\cdot ({\bf r}-{\bf r}^{\,\,\prime}) \; \; 
\frac{H_{1}^{(1)}\left( k_2\,|{\bf r}-{\bf r}^{\,\,\prime}|\right)}
{|{\bf r}-{\bf r}^{\,\,\prime}|} \,\psi^{(2)}({\bf r}^{\,\,\prime}) \right. \nonumber
\end{eqnarray}
\begin{eqnarray}
\centering
-\left. 
H_{0}^{(1)} \left( k_2\,
|{\bf r}-{\bf r}^{\,\,\prime}|
\right)  \frac{\partial \psi^{(2)}({\bf r}^{\,\,\prime})}{\partial \hat n^{\prime}}\right) dS^{\prime}\,.
\label{eq:adentro}
\end{eqnarray}
where the boundary conditions \eqref{eq:contorno-p} and \eqref{eq:contorno-s} must be used to obtain the interior source functions $\psi^{(2)}$ and $\partial \psi^{(2)}/\partial \hat{n}$ in terms of the  
exterior source functions $\psi^{(1)}$ and $\partial \psi^{(1)}/\partial \hat{n}$. 

Knowing the total electromagnetic field allows us to calculate optical characteristics such as the scattering, 
absorption, and extinction cross sections \cite{mishc1,bohren}. 
The time-averaged total power scattered by a two-dimensional particle can be evaluated by calculating the complex Poynting vector flux through an imaginary cylinder of length $L$ and radius $r_{0}$ that encloses the particle 
(see Figure\ \ref{fig:esquema}) 
\begin{equation}
P_{sc}= r_0 L \int_{0}^ {2 \pi} \left\langle{\bf S}_{sc}(r_0,\theta) \right\rangle \cdot {\bf \hat r} \, d\theta,
\label{eq:pot_integral1}
\end{equation}
where 
\begin{equation}
\left\langle{\bf S}_{sc}(r_0,\theta) \right\rangle =\frac{c^2}{8\pi \, \omega \, \eta} {\rm Re\mit} \Big(i{\bf F}(r,\theta) \times \left[\nabla \times {\bf F}(r,\theta)\right]^* \Big),
\label{eq:poynting2}
\end{equation}
and ${\bf F}(r,\theta)=\psi_{sc}(r,\theta)\ {\bf \hat z}$ and $\eta=\mu_1$ ($s$-polarization) or $\eta=\varepsilon_1$ ($p$-polarization). Introducing \eqref{eq:poynting2} into \eqref{eq:pot_integral1}, we obtain
\begin{equation}
P_{sc}= \frac{c^2 \, r_0 \,L}{8\pi \, \omega \, \eta} \int_{0}^ {2 \pi} {\rm Re\mit}  \Big( i \psi_{sc}(r_0,\theta) \, \frac{\partial \psi_{sc}^*}{\partial r}\Big) \, d\theta.
\label{eq:pot_integral2}
\end{equation} 

In the far-field region the calculation of the scattered fields --given by the second term in \eqref{eq:Green2}-- can be greatily simplified using the asymptotic expansion of the Hankel function for large argument. 
After some algebraic manipulation, the following results are found
\begin{eqnarray}
\psi_{sc}(r,\theta)& =& -\,\frac{i \exp ( i \,[  k_1 r-\pi/4] )  } {\sqrt{8 \pi \, k_1\, r}} \,\,F_{ang}(\theta), \label{eq:scatt3}
\end{eqnarray}
\begin{eqnarray}
\frac{\partial \psi^*_{sc}(r,\theta)}{\partial r}& =& i k_1\, \psi_{sc}(r,\theta),
\label{eq:deri_scatt}
\end{eqnarray}
where the angular factor $F_{ang}(\theta)$ is given by 
\begin{align*}
\!\!\!\!\!\!F_{ang}(\theta)=\int_{J(\Gamma)}\Big(ik_1\,[-g'(\tau)cos\theta+
f'(\tau)sin\theta]\psi^{(1)}(\tau)
\end{align*}
\begin{align}
+\frac{\partial \psi^{(1)}}{\partial \hat{n}}(\tau)\Big) 
\,\exp(-i k_1\,[f(\tau)cos\theta+g(\tau)sin\theta]) \,d\tau\,.
\end{align}
When Eqs. (\ref{eq:scatt3}) and (\ref{eq:deri_scatt}) are substituted into 
Eq.  (\ref{eq:pot_integral2}) we get
\begin{eqnarray}
P_{sc}=\frac{c^2 L}{64 \pi^2 \, \omega \, \eta} \int_0^{2 \pi} |F_{ang}(\theta )|^2 \, d\theta.
\label{eq:pot_integral_ang}
\end{eqnarray}
%
%
%
The scattering efficiency $Q_{s}$ is defined as the ratio between the total power scattered by the two-dimensional particle, given by \eqref{eq:pot_integral_ang}, and the incident power $P_{inc}$  intersected by the area $DL$ (see Figure \ref{fig:esquema}). 
Analogously, the absorption efficiency $Q_{a}$ is defined as the ratio between the power $P_{a}$ 
absorbed by the graphene-wrapped two-dimensional particle and $P_{inc}$. 
$P_{a}$ can be obtained as 
%
%
%
\begin{eqnarray}
\!\!\!\!\!\!P_{a}=-\frac{c^2 L}{8 \pi\, \omega \, \eta}
\int_{J(\Gamma)}  {\rm Re\mit}  \big ( {i \, N(\tau) \; 
\psi^{(1)}(\tau) \, \big [\frac{\partial \psi^{(1)}}{\partial \hat n}}(\tau)\big ]^* \big ) \, d\tau 
\label{eq:pot_abs}
\end{eqnarray}
with $N(\tau)=\sqrt{(f{\prime}(\tau))^2+(g{\prime}(\tau))^2}$. 

\section{Results and discussion}\label{sec:results}

\subsection{Wires of circular section}\label{validez}

\begin{figure}[htbp]
\includegraphics[scale=0.33]{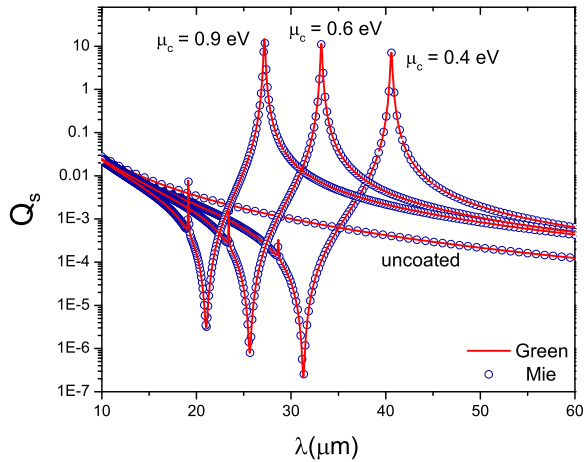}
\caption{Comparison between the Green and the multipolar Mie formalisms results for the p--polarized scattering efficiency $Q_{s}$ of a graphene--coated circular wire with a dielectric core. $R=0.5 \;\mu$m, $\varepsilon_{2}=3.9$, $\mu_{2}=1$,  $\mu_{1}=\varepsilon_{1}=1$,  Kubo parameters for the graphene coating are $T=300^\circ$ K, $\gamma_c=0.1$ meV and $\mu_c=0.4$, $0.6$ and $0.9$ eV. }
\label{Comparacion_dielectrico}
\end{figure}

To assess the validity of the integral formalism sketched in Section \ref{sec:Green} for investigating light scattering and absorption in graphene-coated wires near LSP resonances, we resort to circular geometries where
a reference solution exists \cite{cilindros1,cilindros2,cilindros3,cilindros4}. 
In Figure \ref{Comparacion_dielectrico} we compare the numerical results obtained with the  integral formalism  
described in this paper (solid curves) with the semi-analytical results obtained using Lorenz-–Mie-–Debye solution 
for the scattered fields in the form of infinite series of cylindrical multipole partial waves (circles). The curves 
in this figure represent the scattering efficiency spectra for a circular wire with a radius $R=0.5 \;\mu$m, made with a nonplasmonic, transparent material ($\varepsilon_{2}=3.9$, $\mu_{2}=1$) in a vaccum ($\mu_{1}=\varepsilon_{1}=1$). We used Kubo parameters $T=300^\circ$ K, $\gamma_c=0.1$ meV, different values of $\mu_c$, 
excitation frequencies in the range between $5$ THz (incident wavelength $60 \;\mu m$) and $30$ THz (incident wavelength $10 \;\mu$m) and p--polarized incident waves. The curve corresponding to the uncoated wire, 
not showing any plasmonic feature in this spectral range, is given as a reference. 

An excellent agreement between both formalisms is observed in Figure \ref{Comparacion_dielectrico}.  
The spectral position of the dipolar graphene surface plasmon resonances --at wavelengths near 
$40.60 \;\mu$m ($\mu_c=0.4$ eV), $33.20 \;\mu$m ($\mu_c=0.6$ eV) and $27.16 \;\mu$m ($\mu_c=0.9$ eV)-- 
also agree well with those calculated using  nonretarded analytical expressions \cite{cilindros3}. 
The agreement is also observed for the spectral position of the lower local maxima near $28.65 \;\mu$m (for $\mu_c=0.4$ eV), $23.40 \;\mu$m (for $\mu_c=0.6$ eV) and $19.13 \;\mu$m (for $\mu_c=0.9$ eV), which correspond 
to quadripolar surface current distributions in the graphene coating and are associated with a complex pole in the second coefficient of the multipole expansion (the coefficient of the term that varies twice from positive to negative around the cylinder). 

\begin{figure}[tbp!]
\centering
\includegraphics[scale=0.33]{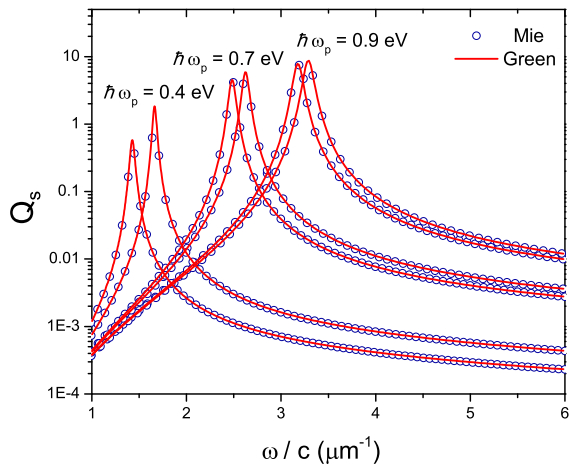}
\caption{\small 
Comparison between the Green and the multipolar Mie formalisms results for the p--polarized scattering efficiency 
$Q_{s}$ of a graphene--coated circular wire with a plasmonic, metallic-like core. $R=50\,$ nm, the 
electric permittivity of the core is described by the Drude model, 
$\varepsilon_2(\omega)=\varepsilon_{\infty}-\omega_p^2/(\omega^2+i\gamma_{m}\omega)$, 
with $\varepsilon_{\infty}=1$, $\gamma_{m}=0.01$ eV and different values of $\hbar \omega_p$ 
(0.4 eV, 0.7 eV and 0.9 eV) and the Kubo parameters for the graphene layer are $T=300^\circ$ K, 
$\gamma_c=0.1$ meV and $\mu_c=0.5\,$ eV.}
\label{Comparacion_metalico}
\end{figure}
%

To further assess the suitability of the Green formalism presented here, we repeat the  comparisons for circular geometries, but with metallic (intrinsically plasmonic) cores instead of dielectric (non plasmonic) cores. 
The results are shown in Figure \ref{Comparacion_metalico}, where we show the spectral dependence of the 
p--polarized scattering efficiency $Q_{s}$ for a graphene-coated metallic wire with $R=50 \;$nm, 
illuminated from vacuum. We assume that the interior electric permittivity is described by the Drude model  
$\varepsilon_2(\omega)=\varepsilon_{\infty}-\omega_p^2/(\omega^2+i\gamma_{m}\omega)$
with $\varepsilon_{\infty}=1$, $\gamma_{m}=0.01$ eV and different values of $\hbar \omega_p$ 
(0.4 eV, 0.7 eV and 0.9 eV).  Kubo parameters for $\sigma(\omega)$ are $T=300^\circ$ K, 
$\gamma_c=0.1$ meV and $\mu_c=0.5$eV. 
We observe that numerical results obtained with the integral formalism (solid curves) and with the Mie solution (circles) are all again in excellent agreement. As explained in \cite{esfera1,cilindros2,cilindros3}, the net effect of the graphene coating is to increase the charge density induced on the surface of the metallic particle, thus blueshifting the resonances of the metallic particle. 

\begin{figure}[tbp!]
\centering
\includegraphics[width=8.8cm]{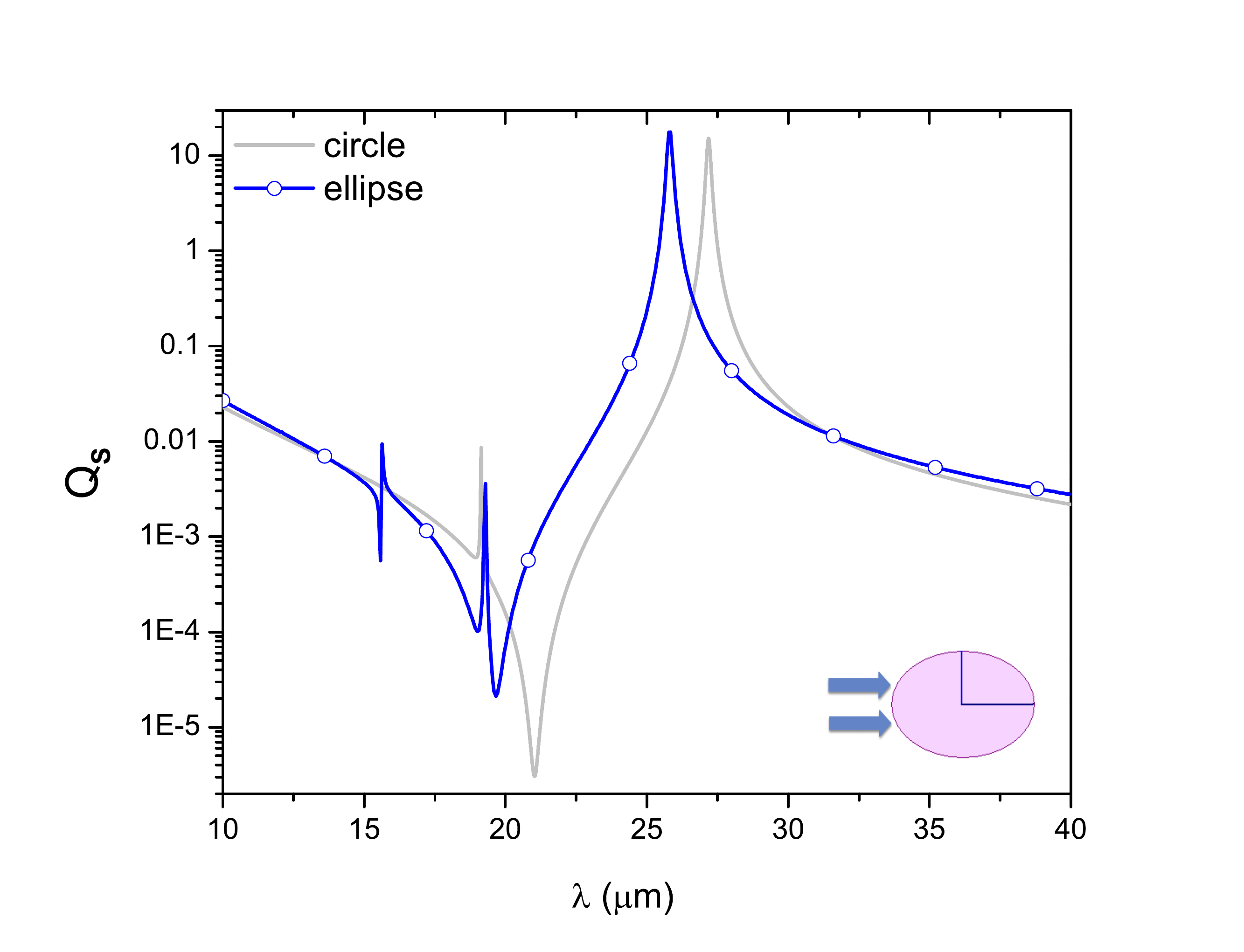}\\
\vspace{-0.5cm}
\includegraphics[width=8.8cm]{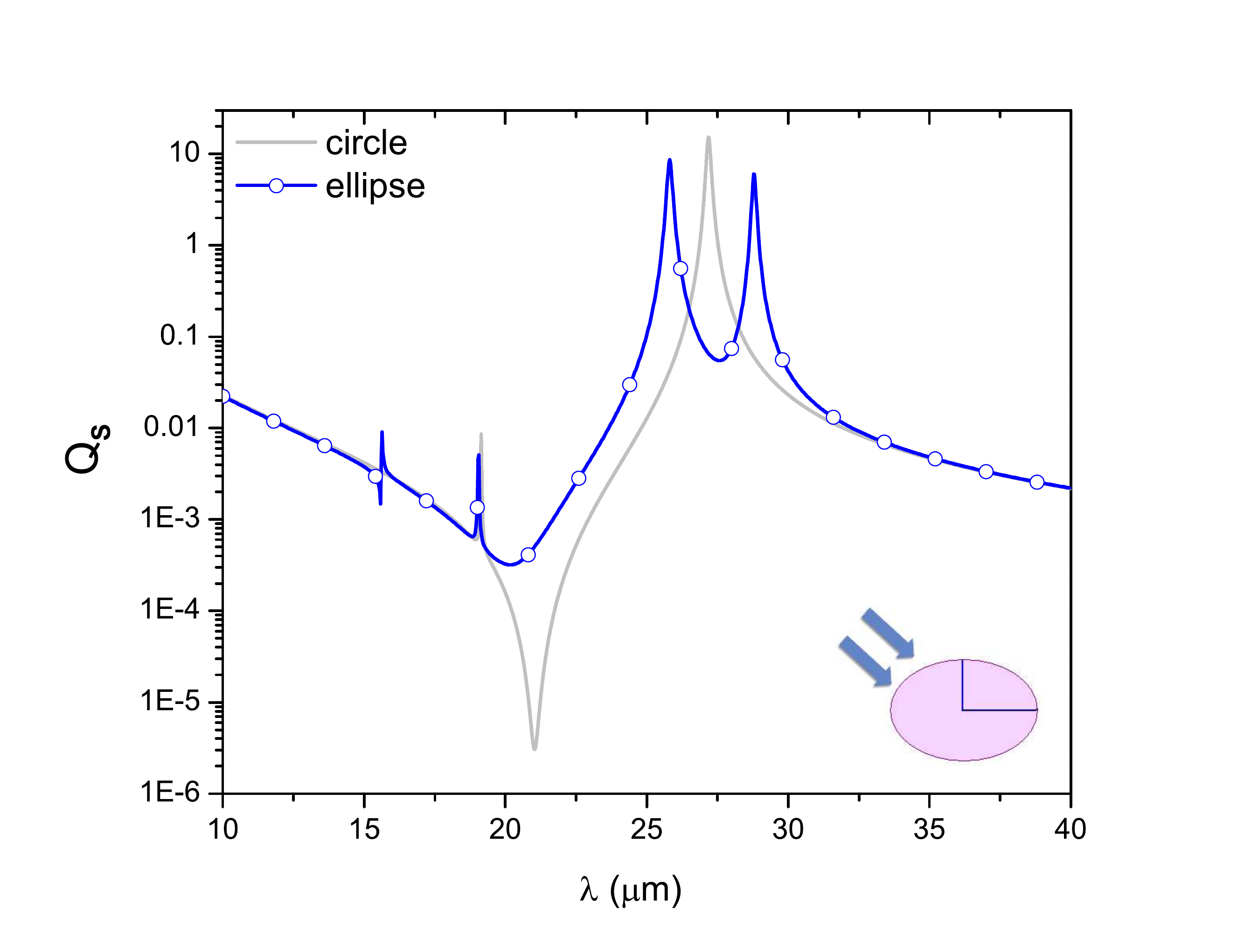}\\
\vspace{-0.5cm}
\includegraphics[width=8.8cm]{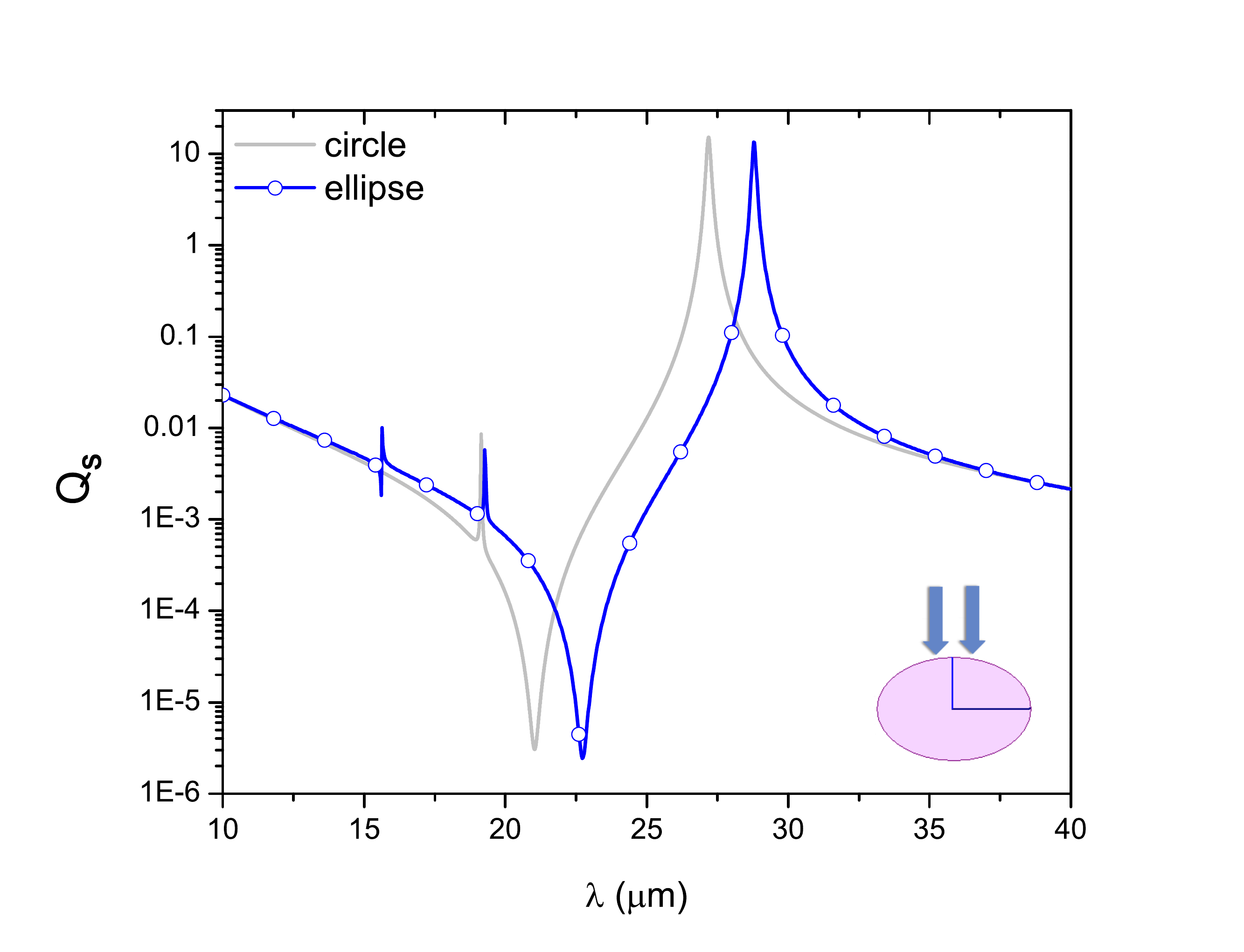}
\caption{\small Scattering efficiency $Q_{s}$  for p--polarized incident waves and for 
illumination direction along (top), at $45^\circ$  (middle) and perpendicular to (bottom) the major axis, 
for a graphene-coated elliptical wire with semi--axes $a=0.55 \;\mu$m and $b=0.45 \;\mu$m, 
$\varepsilon_{2}=3.9$, $\mu_{2}=1$, $\mu_{1}=\varepsilon_{1}=1$  and Kubo parameters $T=300^\circ$ K,  
$\gamma_c=0.1$ meV and $\mu_c=0.9$ eV.  The light gray curve corresponds to a  graphene--coated circular 
wire ($a=b=0.5 \;\mu$m). }
\label{elipse:angulos1}
\end{figure}
%

\subsection{Wires of elliptic section}\label{Elliptic}

Having determined the suitability of the Green approach for the simulation of plasmon resonances in 
graphene--covered wires, we next explore the effects that the departure from circular geometries has on 
the spectrum of graphene LSPs supported by the wire. 
Guided by previous research on metallic nanowires with a nonregular cross section \cite{martin2,martin3}, the complexity of the resonance spectrum is expected to increase when the symmetry of the wire section decreases. 
In order to proceed gradually and keep some degree of control 
to verify the effectiveness of our Green theoretical formulation and numerical codes, 
we consider wires with elliptical section, a geometry which includes the circle as a special case. 

In Figure \ref{elipse:angulos1} we plot the scattering efficiency $Q_{s}$  for p--polarized incident waves and for various angles of incidence for a graphene-coated elliptical wire with major and minor semi--axes 
$a=0.55 \;\mu$m and $b=0.45 \;\mu$m respectively. 
The wire substrate is a nonplasmonic, transparent material ($\varepsilon_{2}=3.9$, $\mu_{2}=1$), the medium of incidence is vaccum ($\mu_{1}=\varepsilon_{1}=1$) and the Kubo parameters are $T=300^\circ$ K, $\gamma_c=0.1$ meV and $\mu_c=0.9$ eV.  
The curve corresponding to a  graphene--coated wire of circular section and $a=b=0.5 \;\mu$m is given as a reference.
We observe that, similar to the case of metallic LSPs \cite{martin2,martin3}, the break of the rotational symmetry 
of the wire section introduces a two-dimensional anisotropy in the angular optical response. 
This anisotropy is particularly evident for the dipolar plasmonic resonance which for the circular case 
($a=b=0.5 \;\mu$m) occurs near $27.16 \;\mu$m and that is split into two different resonant peaks, 
one near $25.81 \;\mu$m  and the other near  $28.78 \;\mu$m 
The first peak corresponds to the illumination direction along the ellipse's major axis 
while the second peak corresponds to the illumination direction perpendicular 
to the major axis, as clearly indicated in Figure \ref{elipse:angulos1} by the 
fact that both resonances are decoupled for illumination directions parallel 
to either of the ellipse's axes and that 
the first (respectively second) peak is absent when the illumination direction is perpendicular to 
(respectively along) the major axis. 
We observe that while the graphene coating always introduces a minimum in the scattering efficiency of 
circular rods --near $21.02 \;\mu$m for the parameters in Figure \ref{elipse:angulos1}, a relevant feature 
in the context of graphene invisibility cloaks \cite{esfera2} and corresponding to a complex zero of the first 
coefficient of the cylindrical multipole expansion \cite{cilindros2}-- the position and magnitude of this minimum depend strongly on the illumination direction, which is another manifestation of two-dimensional anisotropy in the angular optical response of the elliptical wire. 

\begin{figure}[bp!]
\centering
\includegraphics[scale=0.35]{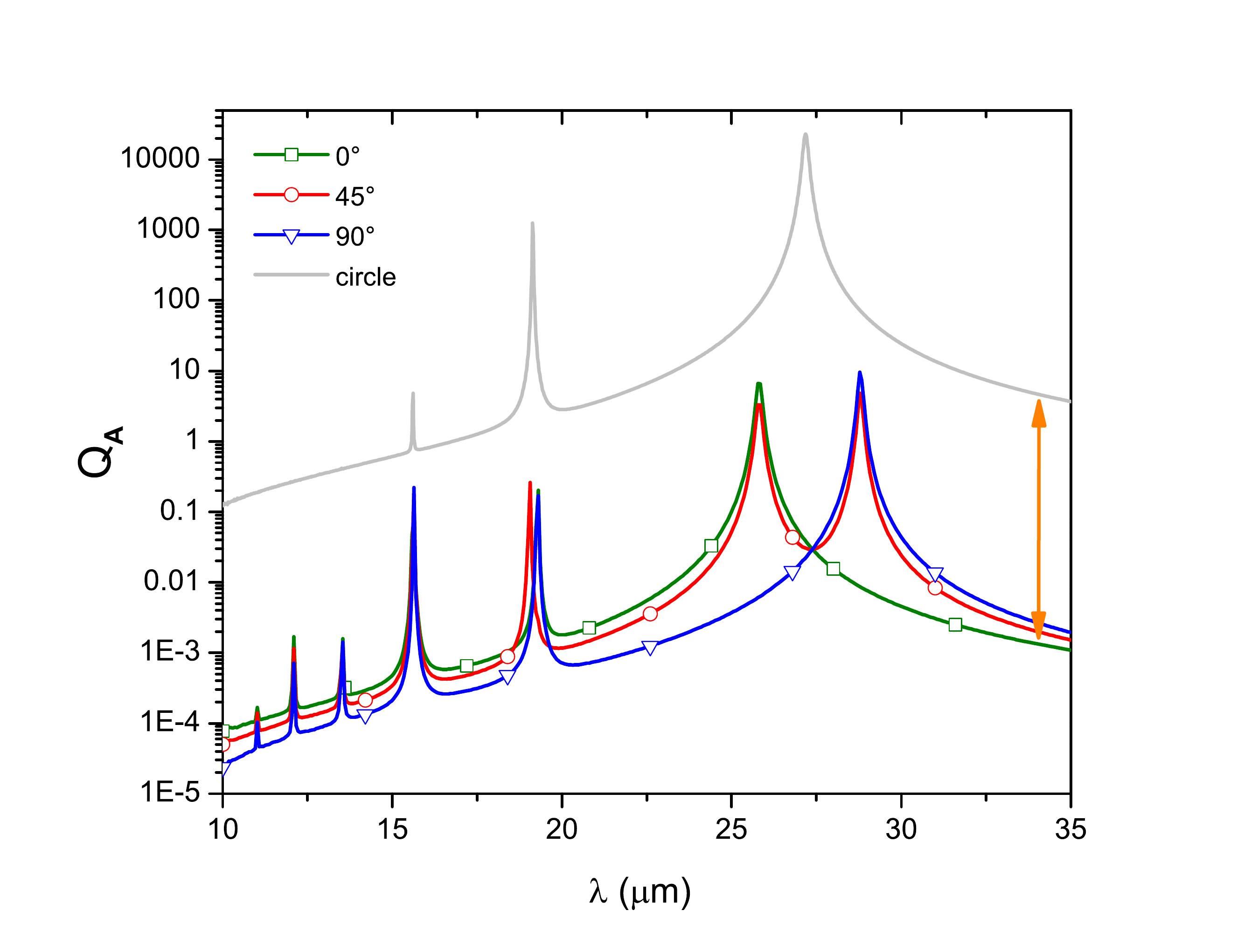}
\caption{\small  Absorption efficiency for different angles of incidence for the same elliptical wire considered in Figure \ref{elipse:angulos1}. 
The illumination direction is: i) along the ellipse's major axis ($0^\circ$); ii) makes an angle  of $45^\circ$  with the ellipse's axes; and iii) perpendicular to the ellipse's  major axis ($90^\circ$). 
The light gray curve corresponds to a  graphene--coated circular 
wire and is displaced, for clarity, by the magnitude indicated by the double-headed arrow.}
\label{figelipse:angulos2}
\end{figure}

Although associated with weaker enhancements of the scattering efficiency, other peaks corresponding to multipolar modes higher than the dipolar mode can also be observed in figure \ref{elipse:angulos1}. 
These higher frequency graphene LSP modes are better appreciated in the near field, as shown in Figure 
\ref{figelipse:angulos2} where we plot absorption efficiency $Q_{a}$  spectra for the same elliptical wire and directions of incidence considered in Figure \ref{elipse:angulos1}. The abosorption spectrum corresponding to the circular case ($a=b=0.5 \;\mu$m) is also given as a reference. 
Apart from the frequency splitting of the  dipolar mode already noted in Figure \ref{elipse:angulos1}, we also observe a splitting in the quadrupolar resonance, which in the circular case occurs at a wavelength near  $19.13 \;\mu$m and that in the elliptical case is split into a peak near $19.27 \;\mu$m,  corresponding to illumination directions parallel to the ellipse's axes, and another peak near $19.06 \;\mu$m,  the only resolved peak when the illumination direction 
makes and angle  of $45^\circ$  with the ellipse's axes. 

\begin{figure}[tbp!]
\centering
\includegraphics[width=9cm]{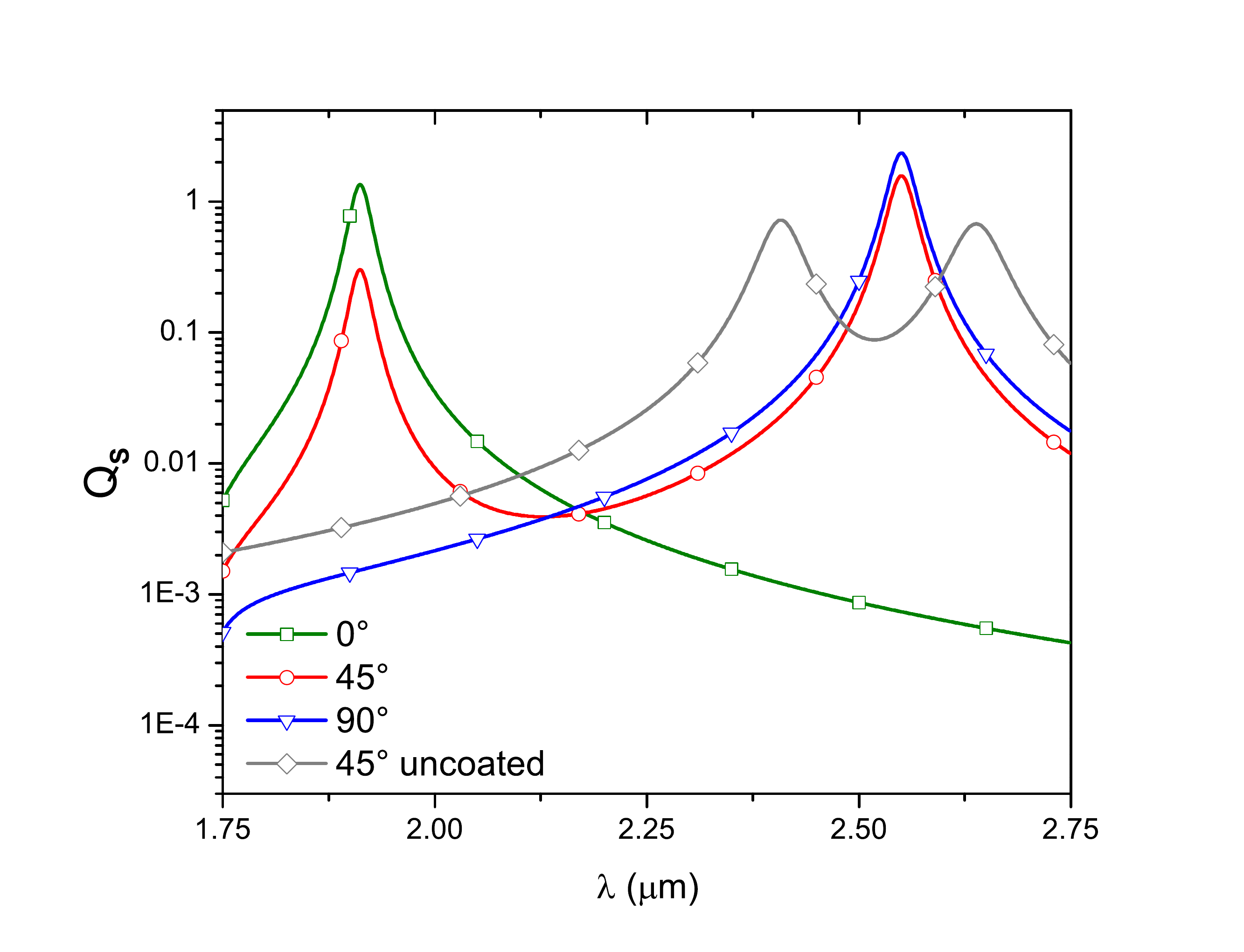} 
\caption{Scattering efficiency spectra for a graphene-coated elliptical wire 
($a=0.030 \;\mu$m and $b=0.025 \;\mu$m) with a plasmonic, metallic core. 
The electric permittivity of the core is described by the Drude model, $\varepsilon_2(\omega)=\varepsilon_{\infty}-\omega_p^2/(\omega^2+i\gamma_{m}\omega)$, with $\varepsilon_{\infty}=1$, $\gamma_{m}=0.01\,$ eV and $\hbar \omega_p=0.7\,$eV. The Kubo parameters for the graphene layer are $T=300^\circ$ K, $\gamma_c=0.1$ meV and $\mu_c=0.7\,$ eV. The illumination direction is: i) along the ellipse's major axis ($0^\circ$); ii) makes an angle  of $45^\circ$  with the ellipse's axes; and iii) perpendicular to the ellipse's  major axis ($90^\circ$). The curve corresponding to the bare elliptical wire illuminated at angle  of $45^\circ$  with the ellipse's axes is given as a reference. }
\label{elipse:metalscat}
\end{figure}
In agreement with the results shown in Figure \ref{Comparacion_metalico} for graphene--coated metallic rods of circular section, in the elliptical case the enhancements of the scattering efficiency are significant only for the dipolar mode. 
This is shown in Figure \ref{elipse:metalscat}, where we observe that the dipolar resonance is now split  
into two peaks corresponding to illumination directions perpendicular and parallel 
to the ellipse's major axis. The curve corresponding to the bare elliptical wire illuminated at an angle of $45^\circ$  with the ellipse's axes is given as a reference. We note that both dipolar resonances are blueshifted compared to those 
of the bare wire, in complete agreement with the fact that the net effect of the graphene coating is to increase the induced charge  density on the cylindrical surface of the metallic wire \cite{cilindros2,cilindros3}. 

\begin{figure}[tbp!]
\centering
\includegraphics[width=9cm]{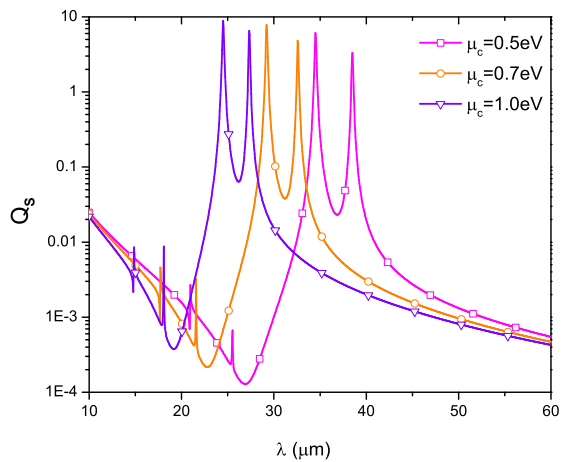} 
\caption{\small   Scattering efficiency $Q_{s}$  for an elliptical wire with three different values of the 
chemical potential $\mu_c$. The illumination direction makes an angle of $45^\circ$  with the major axis of the ellipse, 
the other constitutive and geometrical parameters are the same as those considered in Figure \ref{figelipse:angulos2}.}
\label{elipse:muquimicos}
\end{figure}

Graphene-coated plasmonic particles are particularly attractive because their scattering and absorption characteristics can be tuned by varying the graphene chemical potential $\mu_q$ with the help of a constant electric field (electric field effect, gate voltage), and not only by changing their size or their dielectric constant, as is the case for metallic particles. 
To illustrate this tunability for elliptical wires, we show in Figure \ref{elipse:muquimicos} 
the scattering efficiency $Q_{s}$  for the wire considered in Figure \ref{figelipse:angulos2} and three different values 
of the chemical potential, $\mu_c=0.5\,$ eV,  0.7 eV and 1.0  eV. 
The illumination direction makes an angle of $45^\circ$  with the major axis of the ellipse and the incident wave is p--polarized. We observe that the value of the chemical potential influences the position of the multipolar resonances and the magnitude of the splitting. 

\begin{figure}[tbp!]
\centering
\includegraphics[width=9cm]{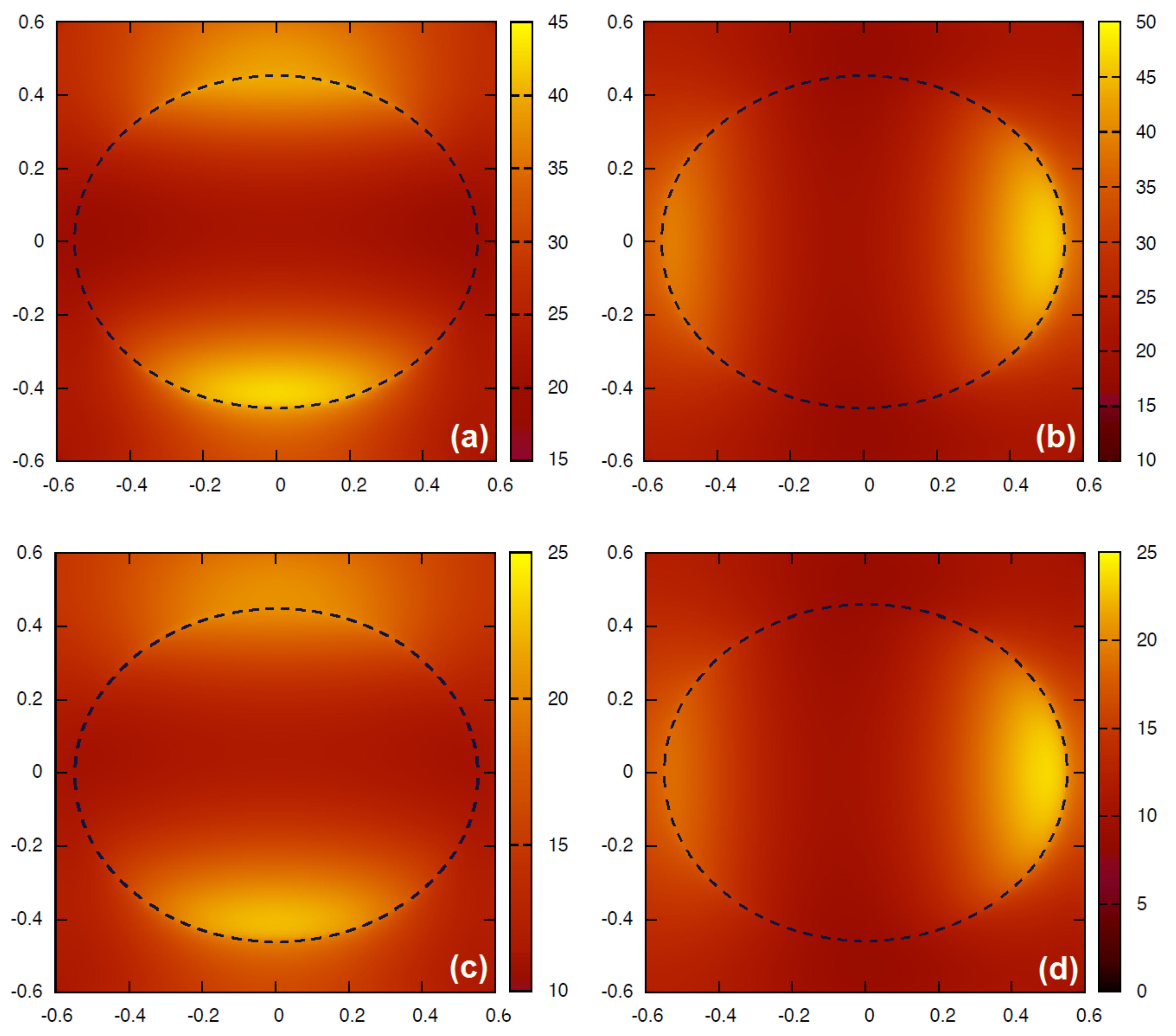}\hspace{-0.5cm}
\caption{\small Map of the near field $|\vec E (x, y)|$ for the graphene-coated elliptical cylinder considered in Figures \ref{elipse:angulos1} and \ref{figelipse:angulos2}. 
a) $\lambda=25.81 \;\mu$m, illumination direction along the ellipse's major axis ($0^\circ$); 
b) $\lambda=28.78 \;\mu$m, illumination direction along the ellipse's minor axis ($90^\circ$); 
c) $\lambda=25.81 \;\mu$m, illumination direction at $45^\circ$  with the ellipse's axes; 
d) $\lambda=28.78 \;\mu$m, illumination direction at $45^\circ$  with the ellipse's axes. }
\label{near:dipolar1}
\end{figure}

\begin{figure}[t!]
\centering
\includegraphics[width=9cm]{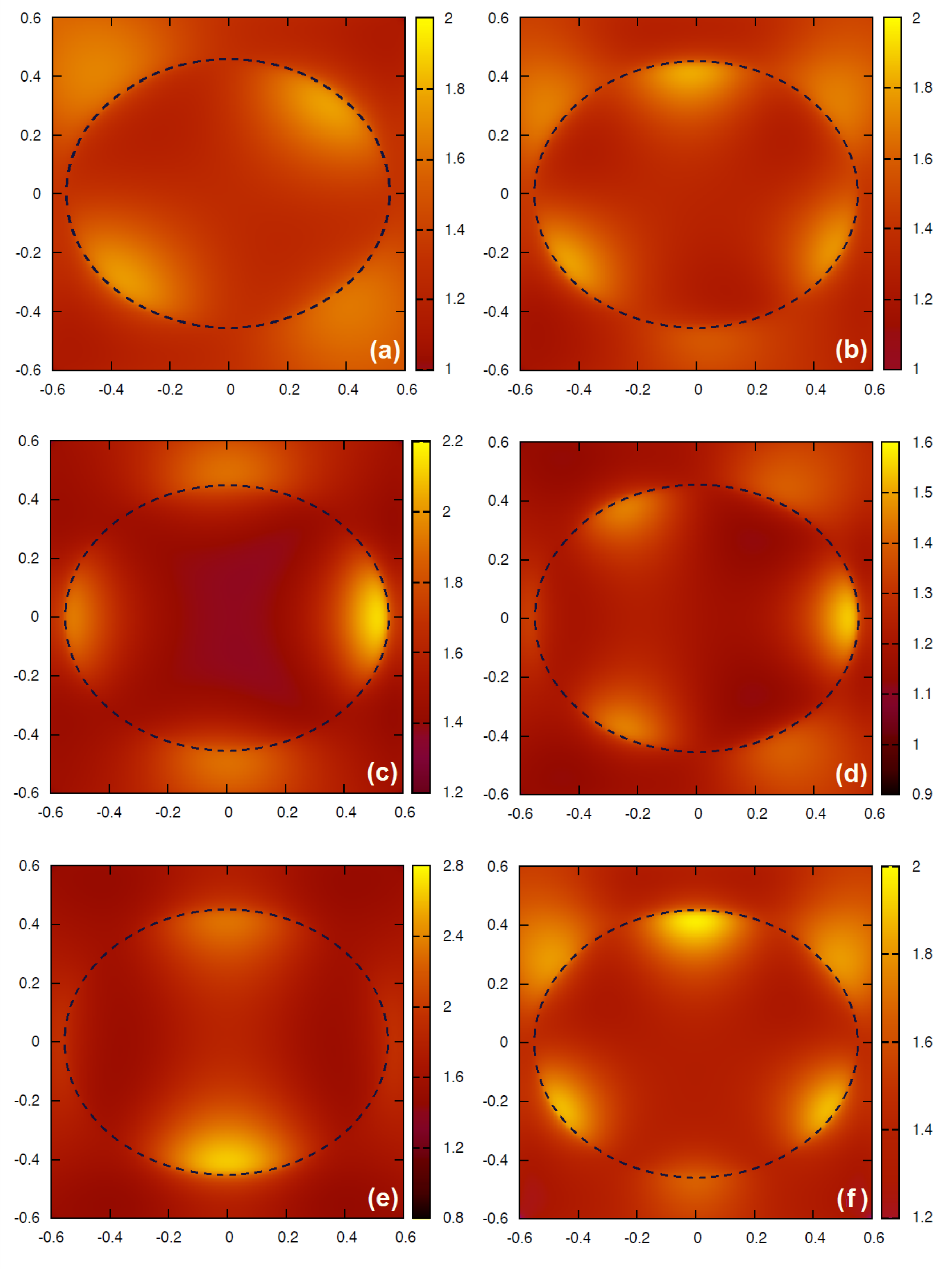}\hspace{-0.5cm}
\caption{\small Map of the near field $|\vec E (x, y)|$ for the graphene-coated elliptical cylinder considered in Figures \ref{elipse:angulos1} and \ref{figelipse:angulos2}. 
a) $\lambda=19.06 \;\mu$m, illumination direction at $45^\circ$  with the ellipse's axes;  
b) $\lambda=15.64 \;\mu$m, illumination direction at $45^\circ$  with the ellipse's axes; 
c) $\lambda=19.27 \;\mu$m, illumination direction along the ellipse's minor axis ($90^\circ$);
d) $\lambda=15.64 \;\mu$m, illumination direction along the ellipse's minor axis ($90^\circ$); 
e) $\lambda=19.27 \;\mu$m,  illumination direction along the ellipse's major axis ($0^\circ$); 
f) $\lambda=15.64 \;\mu$m,  illumination direction along the ellipse's major axis ($0^\circ$). }
\label{near:quadyhexa1}
\end{figure}

In Figure~\ref{near:dipolar1} we plot the spatial distribution of the electric field normalized to the incident amplitude  for the wire considered in Figures \ref{elipse:angulos1} and \ref{figelipse:angulos2}. In Figure~\ref{near:dipolar1}a the incident wavelength is 
$\lambda=25.81 \;\mu$m and the illumination direction is along the ellipse's major axis ($0^\circ$) whereas 
in Figure~\ref{near:dipolar1}b the incident wavelength is $\lambda=28.78 \;\mu$m and the illumination direction is 
along the ellipse's minor axis ($90^\circ$). We observe that in both cases the wire is behaving as an oscillating electric dipole oriented along the direction of the incident field, that is, along the ellipse's  minor axis in Figure~\ref{near:dipolar1}a or 
along the ellipse's major axis in Figure~\ref{near:dipolar1}b. 
Figure \ref{near:dipolar1}c, corresponding to an incident wavelength $\lambda=25.81 \;\mu$m and illumination direction 
making  an angle of $45^\circ$  with the ellipse's axes, shows that in these conditions the electric field inside the wire is parallel to the minor axis, although the components  of the incident field along both ellipse's axes have equal magnitude.  
Figure \ref{near:dipolar1}d, corresponding to the incident wavelength $\lambda=28.78 \;\mu$m and illumination direction 
at $45^\circ$  with the ellipse's axes, shows a similar behavior, except that now the electric field inside the wire is parallel to the major axis, despite the fact that the components of the incident field along both ellipse's axes have equal magnitude. 
The results in Figures \ref{near:dipolar1}c and \ref{near:dipolar1}d are a clear confirmation that 
the values $\lambda=25.81 \;\mu$m and $\lambda=28.78 \;\mu$m correspond to the splitting 
of the degenerate dipolar resonance of the circular wire, as suggested in Figures \ref{elipse:angulos1} and \ref{figelipse:angulos2} by the correspondence between the scattering and absorption efficiency 
spectra for low eccentricity elliptical and circular wires. 
%
%

The agreement between the multipolar order revealed by the topology of the near--field on the one hand and that suggested by 
the correspondence between spectra for circular and for low--eccentricity elliptical wires on the other hand is also observed for 
higher frequency peaks. This is illustrated by the panel in Figure~\ref{near:quadyhexa1}, showing near--field  maps for three illumination directions: at $45^\circ$  with the ellipse's axes (top row), 
along the ellipse's minor axis (middle row) 
and along the ellipse's major axis (bottom row). 
The near--field maps in the left column correspond to the quadrupolar resonance,  in the circular case ($a=b=0.5 \;\mu$m)  located near $19.13 \;\mu$m and split into two close resonant peaks, one at 
$\lambda=19.27 \;\mu$m, associated with field enhancements near vertices and co--vertices, and the other at 
$\lambda=19.06 \;\mu$m, associated with field enhancements near the diagonals of the rectangle circumscribed 
to the ellipse. 
The maps in the right column correspond to hexapolar resonances near  $\lambda=15.64 \;\mu$m (near 
$15.61 \;\mu$m in the circular case). Although the topology of the hexapolar near--field 
when the illumination direction is at $45^\circ$  with the ellipse's axes is very different to the topology 
when the illumination direction is along the axes, symmetry--breaking splittings of the hexapolar peaks are not 
resolvable in the scattering and absorption efficiency spectra. 

\section{Summary and conclusions}
\label{sec:final}

In conclusion, we have presented in this paper an electromagnetically rigorous  method based on Green's second identity for studying the plasmonic response of graphene--coated wires of arbitrary shape. 
To validate the method, we compare the numerically computed scattering and absorption efficiencies 
in the particular case of graphene--coated wires of circular section with the results obtained from a 
multipolar Mie theory. All the results agree excellently, both for metallic and dielectric substrates. 
To explore the effects that the break of the rotational symmetry of the wire section has in the plasmonic features of the scattering and absorption response, we apply the Green formulation to the case of graphene-coated wires of elliptical section. 
Compared with the scattering and absorption efficiency spectra for graphene--coated wires of circular section, 
and as it might be expected on symmetry grounds, a frequency splitting of multipolar plasmonic resonances is 
observed in the spectra for low--eccentricity elliptical wires. To illustrate the application of the Green method 
in the near--field we investigate the spatial distribution of the electromagnetic field near the graphene coating for different frequencies and directions of incidence.  
As a further test on the validity of the Green formulation, we show that the multipolar order revealed by the 
topology of the near field agrees perfectly well with the multipolar order obtained from 
the correspondence between spectra for circular and for low--eccentricity elliptical wires. 
The method presented here should be useful in the design and engineering of graphene--wrapped particles 
with tailored properties for specific plasmonic applications, including photovoltaic devices, nanoantennas, 
switching, biosensing and even medical treatments.

\vspace{0,8cm}
 
\noindent Funding. Consejo Nacional de Investigaciones Cient\'{\i}ficas y T\'ecnicas (CONICET PIP 1800); Universidad de Buenos Aires (UBA 20020100100327); Universidad Aut\'onoma de Baja California (UABC) and Consejo Nacional de Ciencia y Tecnolog\'ia (CONACYT).


\bigskip

\begin{thebibliography}{1}


\bibitem{geim1} 
A. K. Geim, “Graphene: status and prospects," Science \textbf{324}, 1530-34 (2009).

\bibitem{bonaccorso1}
F. Bonaccorso et al, ``Graphene, related two-dimensional crystals, and hybrid systems for energy conversion and storage," Science \textbf{347}, (6217) (2015).

\bibitem{bonaccorso2}
F. Bonaccorso et al, ``Science and technology roadmap for graphene, related two-dimensional crystals, and hybrid systems," Nanoscale \textbf{7}, 4598-4810 (2015). 

\bibitem{ssc1}
F. M. Kin, J. Long, W. Feng, and T. F. Heinz, ``Optical spectroscopy of graphene: From the far infrared to the ultraviolet," Solid State Commun. \textbf{152}, 1341–49 (2012).

\bibitem{conven1}
M. Grande et al, ``Fabrication of doubly resonant plasmonic nanopatch arrays on graphene," Appl. Phys. Lett. \textbf{102}, 231111 (2013).

\bibitem{conven2}
M. Hashemi, M. H. Farzad, N. A. Mortensen, and S. Xiao, ``Enhanced absorption of graphene in the visible region by use of plasmonic nanostructures," J. Opt. \textbf{15}, 055003 (2013).

\bibitem{sp-graf1}
M. Jablan, M. Soljacic, and H. Buljan, ``Plasmons in Graphene: Fundamental Properties and Potential Applications," Proceedings of the IEEE \textbf{101} (7), 1689-704 (2013). 

\bibitem{sp-graf2}
T. Low and P. Avouris, “Graphene Plasmonics for Terahertz to Mid--Infrared Applications," ACS nano \textbf{8}, (2) 1086-101 (2014). 

\bibitem{esfera2} 
M. Farhat, C. Rockstuhl, and H. Bagci,
``A 3D tunable and multi-frequency graphene plasmonic cloak," Opt. Express \textbf{21}, 12592-603 (2013).

\bibitem{esfera1} 
T. Christensen, A. P. Jauho, M. Wubs, and N. Mortensen, ``Localized plasmons in graphene-coated nanospheres," Phys. Rev. B \textbf{91}, 125414 (2015).

\bibitem{esfera3} 
B. Yang, T. Wu, Y. Yang, and X. Zhang, ``Tunable subwavelength strong absorption by graphene wrapped dielectric particles," J. Opt. \textbf{17}, 035002 (2015).

\bibitem{cil4}
Z. R. Huang et al, ``A mid--infrared fast--tunable graphene ring resonator based on guided--plasmonic wave resonance on a curved graphene surface," J. Opt. \textbf{16}, 105004 (2014). 

\bibitem{cilindroconico} 
T. J. Arruda, A. S. Martinez, and F. A. Pinheiro, ``Electromagnetic energy within coated cylinders at
oblique incidence and applications to graphene coatings," J. Opt. Soc. Am. A \textbf{31}, (2014). 

\bibitem{cilOE} 
J. Zhao et al, ``Surface-plasmon-polariton whispering-gallery mode analysis of the graphene monolayer coated InGaAs nanowire cavity," Opt. Express \textbf{22}, 5754–61 (2014).	

\bibitem{cilindros1} 
R. J. Li, X. Lin, S. S. Lin, X. Liu, and H. S. Chen, ``Tunable deep--subwavelength superscattering using graphene monolayers," Opt. Lett. \textbf{40}, (2015).

\bibitem{cilindros2} 
M. Riso, M. Cuevas, and R. A. Depine, ``Tunable plasmonic enhancement of light scattering and absorption in graphene-coated subwavelength wires," J. Opt. \textbf{17}, 075001 (2015).

\bibitem{cilindros3} 
M. Cuevas, M. Riso, and R. A. Depine, ``Complex frequencies and field distributions of localized surface plasmon modes in graphene-coated subwavelength wires," J. Quant. Spectrosc. Ra. \textbf{173}, (2015).

\bibitem{cilindros4} 
E. Velichko, ``Evaluation of a graphene-covered dielectric microtube as a refractive-index sensor in the terahertz range," J. Opt. \textbf{18}, 035008 (2016).

\bibitem{fabric1} 
N. Kumar and S. Kumbhat, \textit{Essentials in Nanoscience and Nanotechnology} (New York: Wiley, 2016). 

\bibitem{C2gribonexp01}
L. Ju et al, “Graphene plasmonics for tunable terahertz metamaterials," Nat. Nanotechnol. \textbf{6} (10) 630-34 (2011).

\bibitem{MMMM} 
 A. A. Maradudin, T. Michel, A. McGurn, and E. R. Mendez,
“Enhanced backscattering of light from a random grating," Ann. Phys. \textbf{203}, 255-307 (1990).

\bibitem{civ2NL}  C. Valencia, E. M\'endez, and B. Mendoza, ``Second harmonic generation in the scattering of light by two-dimensional particles," J. Opt. Soc. Am. B \textbf{20}, 2150–61 (2003).

\bibitem{martin2}  
J. P. Kottmann, O. J. F. Martin, D. R. Smith, and S. Schultz, ``Plasmon resonances of silver nanowires with a nonregular cross section," Phys. Rev. B \textbf{64}, 235402 (2001).

\bibitem{martin3} 
J. P. Kottmann, O. J. F. Martin, D. R. Smith, and S. Schultz, ``Field polarization and polarization charge distributions in plasmon resonant nanoparticles," 
New J. Phys. \textbf{2}, 27 (2000).

\bibitem{kubo2} 
S. A. Milkhailov, K. Siegler, ``New Electromagnetic Mode in Graphene," Phys. Rev. Lett. \textbf{99}, 016803 (2007). 

\bibitem{kubo1} 
L. A. Falkovsky, ``Optical properties of graphene and IV--VI semiconductors," Phys. Usp. \textbf{51}, 887-897 (2008).

\bibitem{mishc1}
M. Mishchenko, L. D. Travis, and A. A. Lacis, \textit{Scattering, Absorption, And Emission Of Light By Small Particles} (Cambridge: Cambridge University Press, 2002).

\bibitem{bohren}
C. F. Bohren and D. R. Huffman, \textit{Absorption and scattering of light by small particles} (New York: Wiley, 1983). 


\end{thebibliography}
\end{document}